\documentclass[runningheads]{llncs}
\usepackage[T1]{fontenc}
\usepackage{multirow}
\usepackage{amssymb}
\usepackage{amsmath} 
\usepackage{booktabs}
\usepackage[misc]{ifsym}

\usepackage{graphicx}
\begin{document}
\title{DB-SAM: Delving into High Quality Universal Medical Image Segmentation}
%
%
\author{Chao Qin\inst{1}$^{\textrm{(\Letter)}}$, Jiale Cao\inst{2}, Huazhu Fu\inst{3}, Fahad Shahbaz Khan\inst{1, 4}, \\ Rao Muhammad Anwer\inst{1}}


%
\authorrunning{C. Qin et al.}
%
\institute{Mohamed bin Zayed University of Artificial Intelligence, Abu Dhabi, United Arab Emirates \\ \email{chao.qin@mbzuai.ac.ae} \\ \and Tianjin University, Tianjin, China \and Institute of High Performance Computing, Agency for Science, Technology and Research, Singapore \and 
Linköping University, Linköping, Sweden}
\maketitle              
\begin{abstract}
Recently, the Segment Anything Model (SAM) has demonstrated promising  segmentation capabilities in a variety of downstream segmentation tasks. However in the context of universal medical image segmentation there exists a notable performance discrepancy when directly applying SAM due to the domain gap between natural and 2D/3D medical data.  In this work, we propose a dual-branch adapted SAM framework, named DB-SAM, that strives to effectively bridge this domain gap. Our dual-branch adapted SAM contains two  branches in parallel: a ViT branch and a convolution branch. The ViT branch incorporates a learnable channel attention block after each frozen attention block, which captures domain-specific local features. On the other hand, the convolution branch employs a light-weight convolutional block to extract domain-specific shallow features from the input medical image. To perform cross-branch feature fusion, we design a bilateral cross-attention block and a ViT convolution fusion block, which dynamically combine diverse information of two branches for mask decoder. Extensive experiments on large-scale medical image dataset with various 3D and 2D medical segmentation tasks reveal the merits of our proposed contributions. On 21 3D medical image segmentation tasks, our proposed DB-SAM  achieves an absolute gain of 8.8\%, compared to a recent medical SAM adapter in the literature. The code and model are available at https://github.com/AlfredQin/DB-SAM.

\end{abstract}
\section{Introduction}
Medical image segmentation plays a pivotal role in various clinical applications, including disease  diagnosis and surgical planning. In past years, deep learning-based methods have achieved significant progress on medical image segmentation tasks. Most existing methods \cite{zhou2018unet++,swinunet,chen2021transunet} are typically task-specific in that they usually focus on a specific segmentation task, for which the dataset is captured  through a specific type of medical device. As a result, these methods have limited generalization ability to different medical segmentation tasks, especially when the  medical device belongs to  different types. 
Recently, a class-agnostic segmentation model, named SAM \cite{kirillov2023segment}, has been introduced to perform universal segmentation in natural images. 
SAM is trained on large-scale data and exhibits impressive generalization ability on various down-stream segmentation tasks.
However, its segmentation quality deteriorates when directly adapting to 2D and 3D medical image segmentation. This is due to a large domain gap between natural images and medical images. 
A straightforward way to adapt SAM for universal medical image segmentation is re-training the entire SAM model on medical image datasets. However, this is challenging since it requires significant computational resources along with the availability of a very large-scale medical image dataset. Compared to natural images, medical images are relatively scarce and expensive. To address this issue, the recently introduced MedSAM \cite{ma2024segment} focuses on only fine-tuning the mask decoder of SAM, achieving better universal medical image segmentation performance compared to the vanilla SAM. Nonetheless, this fine-tuning strategy does not fully harness domain-specific (medical) knowledge and likely results in sub-optimal segmentation performance, especially on organs with intricate outlines. In contrast, the concept of adapter that is prevalent in natural language processing (NLP)  provides a potential avenue. The adapter module is usually parameter-efficient and can be easily integrated into a pre-trained model. During training, only the adapter module is trained and the pre-trained model parameters are frozen. 
In this work, we propose a dual-branch framework, named DB-SAM, that adapts SAM for high-quality universal medical image segmentation. We introduce a novel dual-branch encoder comprising \textbf{(i)} a ViT branch with inserted channel attention blocks for domain-specific locality inductive bias, and \textbf{(ii)} a convolution branch equipped with lightweight convolutional blocks for extracting shallow features. A bilateral cross-attention block is designed for effective cross-layer feature fusion between ViT branch and convolution branch. Finally, the output features of two branches are fused through an automatic selective mechanism.
Our proposed DB-SAM is evaluated on a comprehensive dataset drawn from 30 public medical datasets covering both 3D and 2D  images of different modalities. Our experimental results demonstrate the effectiveness of the proposed DB-SAM, leading to consistent improvement in performance on a diverse set of 2D and 3D medical segmentation tasks. In case of 3D medical image segmentation, the proposed DB-SAM achieves an absolute gain of $6\%$ and $8.9\%$ in terms of DSC and NSD, respectively, compared to the baseline MedSAM \cite{ma2024segment}.

\section{Method}
\begin{figure*}[t!]
  \centering
  \includegraphics[width=1.0\linewidth]{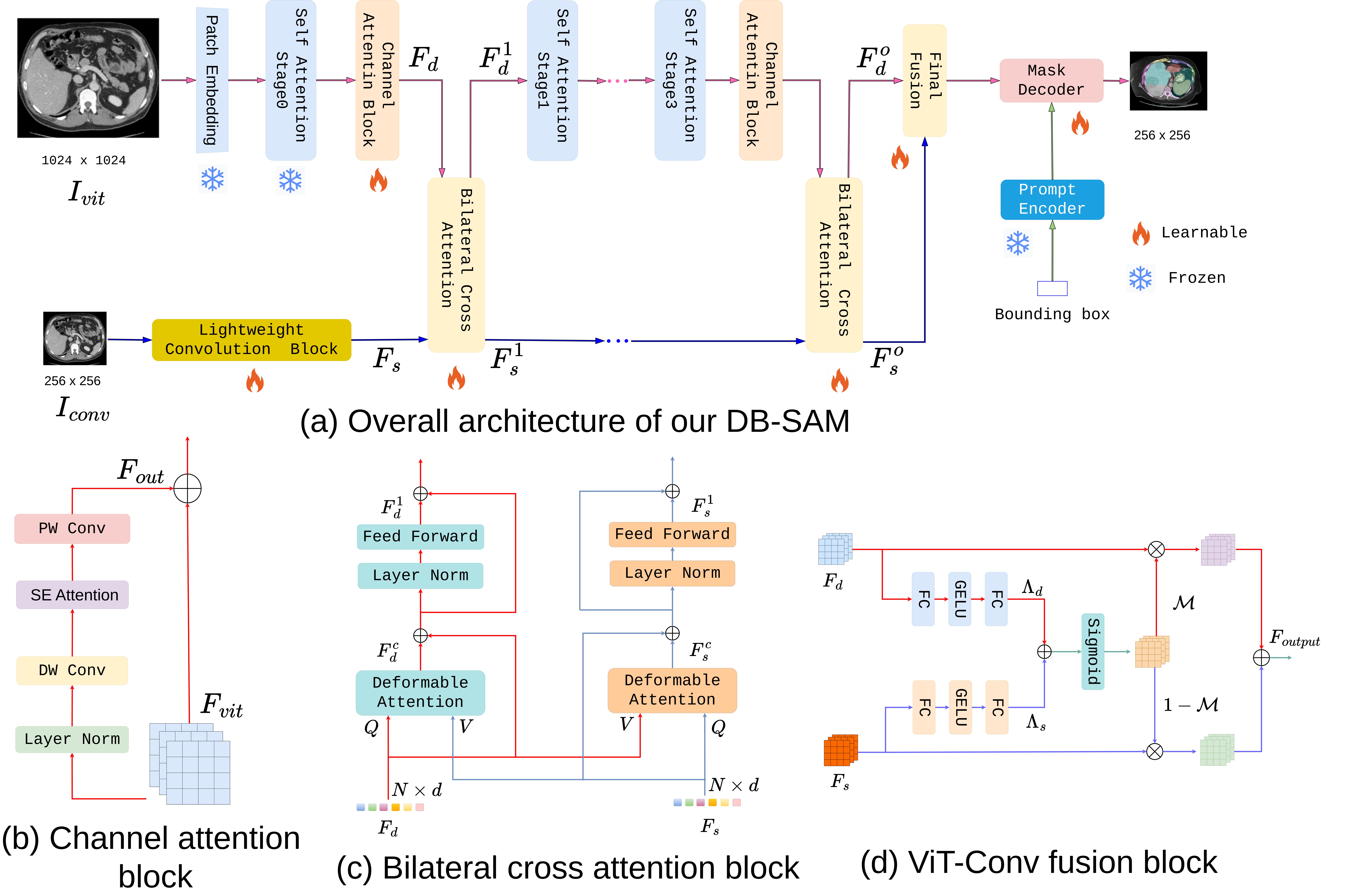}
  \caption{\textbf{(a)} Overall architecture of our DB-SAM. Our DB-SAM contains two branches: one ViT branch and one convolution branch. The ViT branch incorporates channel attention block \textbf{(b)} to capture domain-specific high-level features, while the convolution branch adopts light-weight convolution blocks to extract shallow features. For cross-branch fusion, we introduce a bilateral cross-attention operation \textbf{(c)} and ViT-Conv fusion module \textbf{(d)} to adaptively combine the features. Finally, the fused features and prompt embeddings are fed to mask decoder.}
  \label{fig:model}
\end{figure*}

Fig.~\ref{fig:model}(a) shows overall architecture of our  method. Given an input image $I$, we first resize the input image into two images of different resolutions that are represented as $I_{vit}\in \mathbb{R}^{1024 \times 1024}$ and $I_{conv}\in \mathbb{R}^{256 \times 256}$. These two images are fed to two different branches of our dual-branch image encoder, where the image $I_{vit}$ goes through the ViT branch and the image $I_{conv}$ goes through the convolution branch. The ViT branch comprises a patch embedding block, multiple attention blocks, and multiple channel attention blocks. The convolution branch contains multiple light-weight convolutional blocks. To fuse the features between two branches, the bilateral cross-attention block is designed to perform cross-layer feature fusion. The output features of two branches are represented as $F_{d}^o$ and $F_{s}^o$, which are fused as final image embeddings for mask decoder. In addition, we convert the bounding-box into prompt embeddings using a prompt encoder. Based on the image embeddings  and prompt embeddings, we employ mask decoder to predict corresponding segmentation map.

\subsection{ViT Branch}
The SAM adopts the transformer ViT as image encoder, and trains it on large-scale natural image dataset. To keep the strong feature representation ability of pre-trained ViT, we do not fine-tune the weights of pre-trained ViT during training, and introduce a local adapter module to introduce a locality inductive bias. Inspired by the  advances in EfficientNetV2 \cite{tan2021efficientnetv2}, our local adapter module contains a channel attention block attached after the attention block of ViT. Different to attention block,  the channel attention block is learnable during training and aims to extract high-level domain-specific features from the different levels of ViT encoder. As a result, our ViT branch adapts the original ViT  to medical image segmentation without losing the strong feature representation ability.

Fig.~\ref{fig:model}(b) shows the structure of channel attention block, which is written as
\begin{equation}
F_{\text{out}} = F_{\text{vit}} + \text{Conv}_{1 \times 1}\left(\text{SE}\left(\text{DWConv}_{3 \times 3}\left(\text{LN}(F_{\text{vit}})\right)\right)\right)
\end{equation}
where \( F_{\text{vit}} \) represents input embeddings from ViT attention block, \( \text{LN} \) represents  layer normalization, \( \text{DWConv}_{3 \times 3} \) denotes depth-wise $3\times 3$ convolution, \( \text{SE} \) refers to the squeeze and excitation block~\cite{hu2018squeeze}, and \( \text{Conv}_{1 \times 1} \) is the point-wise convolution. The channel attention block is a simple and efficient operation, which does not introduce much computational cost for  ViT encoder.

\subsection{Convolution Branch}
The original ViT  employs the  patch embeddings to  downsample the input image $16$ times and then extracts deep  features. We argue that the downsampling  is not good to keep locality inductive bias. To incorporate more local details for accurate segmentation, we introduce a  convolution branch that employs light-weight convolution blocks to directly extract shallow features from resized image $I_{conv}$. In addition,  we introduce a bilateral cross attention block to fuse deep features from ViT branch and shallow features from convolution branch,

As  in Fig.~\ref{fig:model}, we first feed the resized  image $I_{conv}$ to  the light-weight convolution block to extract shallow features $F_{s}$. Afterwards, we employ the bilateral cross attention block to perform feature fusion. The generated features for two branches are represented as shallow features $F_{s}^{1}$ and deep features $F_{d}^{1}$. Then we feed the features $F_{d}^1$ to next self-attention blocks in the ViT branch. Similarly, we perform feature fusion of output features of each block. The final features for two branches are represented as  $F_{s}^o$ and  $F_{d}^o$. We introduce the light-weight convolution block and bilateral cross attention block in details as follow.

\noindent\textbf{Light-weight convolution block:}  Our light-weight convolution block consists of two $3\times3$  and three $1\times1$ convolutional layers. We add a batch normalization layer and a ReLU layer after convolutional layer at the first four layers.

\noindent\textbf{Bilateral cross attention block:} Fig.~\ref{fig:model}(c) shows the structure of bilateral cross attention block that aims to perform cross-branch feature fusion. Assuming that the deep feature is $F_{d}$ and the shallow feature is $F_{s}$, we employ deformable attention \cite{zhu2020deformable}  to fuse them as in \cite{chen2022vision}. For the ViT branch, we treat deep feature $F_{d}$ as query, and treat shallow feature $F_{s}$ as  key and value. For the convolution branch, we treat shallow feature $F_{s}$ as query, and treat deep feature $F_{d}$ as  key and value. The features generated by deformable attention are respectively represented as $F_{d}^{c}$ and $F_{s}^{c}$, which are fed to a layer normalization layer and a feed-forward layer to generate the output features $F_{d}^{1}$ and $F_{s}^{1}$. The feed-forward layer contains two MLP layers and a GeLU layer. In addition, we add the residual connection in both deformable attention and feed forward layers.

\subsection{ViT-Conv Fusion Block}
We  design a fusion module with an automatic selective mechanism to integrate the diverse information from ViT branch and  convolution branch. 
Fig.~\ref{fig:model}(d) gives the architecture.
The feature from a given branch, denoted as \( F_{d} \) or \( F_{s} \), is initially processed through a channel attention layer. This layer comprises a  fully connected squeeze layer, a GELU activation layer, and a fully connected restore layer. This process yields the logits \( \Lambda_{d} \) or \( \Lambda_{s} \). The output logits from each branch's channel attention layer are then combined, leading to the creation of an element-wise selective mask. This mask is derived by applying the sigmoid  to the summed features as
 $\mathcal{M} = \text{Sigmoid}(\Lambda_{d} + \Lambda_{s})$.   
The final fusion output, representing an amalgamation of both branches, is computed as follows:
\begin{equation}
F_{\text{output}} = F_{d}^o \otimes \mathcal{M} + F_{s}^o \otimes (1 - \mathcal{M}),   
\end{equation}
where \( F_{d}^o \) and \( F_{s}^o \) represent the features from the ViT and convolution branches, respectively, while \( \otimes \) signifies element-wise multiplication. Each token in the final output feature map learns both the global context information from the ViT branch and the local spatial information from the convolution branch. This is achieved by adaptive fusion between the two branches, ensuring a  comprehensive representation of the features.

\begin{table}[!t]
\centering
\footnotesize
\caption{Performance comparison between our method, MedSAM, and SAM on 21 3D medical image segmentation tasks evaluated by DSC and NSD. Our model achieves significant and consistent improvements across all the tasks. Best results are in bold.}
\resizebox{\linewidth}{!}{
\begin{tabular}{|cc|ccc|ccc|}
\hline
\multirow{2}{*}{Segmentation Task}     &   \multirow{2}{*}{Modality}       & \multicolumn{3}{c|}{DSC (\%)}                & \multicolumn{3}{c|}{NSD (\%)}                               \\ \cline{3-8} 
 &     & SAM~\cite{kirillov2023segment}        & MedSAM~\cite{ma2024segment} & \textbf{DB-SAM (Ours)}   & SAM~\cite{kirillov2023segment}         & MedSAM~\cite{ma2024segment} & \textbf{DB-SAM (Ours)}  \\ \hline
Brain Ventricles  & MR-T1     & 41.96  & 74.82   & \textbf{78.95}        & 31.26     & 78.17                          & \textbf{82.08} \\
Brain Ventricles  & MR-T2     & 39.56  & 72.87    & \textbf{77.02}  & 31.39         & 75.01     & \textbf{81.23}                      \\
Brain Tumor       & MR-FLAIR  & 74     & 89.15        & \textbf{92.49} & 38.27          & 76.13   & \textbf{85.77}                     \\
Cerebellum        & MR-T1     & 83.25 & 93.31   & \textbf{94.47}                      & 44.55         & 82.39     & \textbf{88.32}\\
Cerebellum        & MR-T2    & 81.88 & 90.78     & \textbf{92.69}  & 38.84            & 70.01                     & \textbf{76.77}   \\
Gallbladder       & MR       & 61.97 & 77.78   & \textbf{87.43}     & 36.34     & 76.36                      & \textbf{84.57}   \\
Left Ventricle    & MR       & 68.44 & 88.91    & \textbf{91.15}          & 55.73   & 91.05                     & \textbf{94.10} \\
Right Ventricle   & MR       & 72.11  & 85.92  & \textbf{90.51}       & 68.98    & 88.85                      & \textbf{94.97}  \\
Liver             & MR       & 80.38  & 93.9   & \textbf{96.17}         & 33     & 80.13                      & \textbf{89.87}  \\
Pancreas          & MR       & 51.11  & 80.07   & \textbf{85.16}    & 32.71       & 79.42    & \textbf{88.26}                      \\
Prostate          & MR-ADC   & 79.61 & 92.25    & \textbf{93.25}      & 60.12     & 92.72                       & \textbf{94.13}  \\
Prostate          & MR-T2    & 79.39  & 92.18   & \textbf{93.45}   & 57.6          & 92   & \textbf{94.32}                       \\
Abdomen Tumor     & CT       & 42.86  & 65.54 & \textbf{78.31}  & 34.49         & 64.99                  & \textbf{79.62}       \\
Gallbladder       & CT       & 47.28 & 84.36    & \textbf{90.12}   & 30.48           & 87.07   & \textbf{94.92}                  \\
Head-Neck Tumor   & CT       & 23.87 & 68.29    & \textbf{76.14}    & 23.88           & 47.36               & \textbf{56.27}      \\
Liver             & CT       & 74.21    & 91.42   & \textbf{96.21}         & 26.08    & 76.3                      & \textbf{91.35} \\
Lung Infections   & CT       & 32.54  & 60.01    & \textbf{78.51}    & 25.84          & 58.89                  & \textbf{78.98}      \\
Pancreas          & CT       & 43.53   & 76.76   & \textbf{83.8}         & 32.38   & 81.09                      & \textbf{89.45}  \\
Pleural Effusion  & CT       & 9.52 & 59.46   & \textbf{75.52}    & 11.19         & 75.77    & \textbf{90.96}                   \\
Stomach           & CT       & 68.95 & 82.66  & \textbf{92.54}      & 35.49     & 70.93                       & \textbf{87.74}  \\
Head-Neck Tumor   & PET      & 72.45 & 81.17  & \textbf{84.12}      & 38.57      & 62.76                        & \textbf{67.78}   \\ \hline
Average           &          & 58.52  & 81.02  & \textbf{87.05}  & 37.49           & 76.54                   & \textbf{85.31}    \\ \hline
\end{tabular}%
}
\label{tab:comparision on 3D tasks}
\end{table}

\section{Experiments}


\begin{table}[!t]
\centering
\footnotesize
\caption{Performance comparison between our method and MedSAM as well as SAM on 9 2D medical image segmentation tasks evaluated by Normalized Surface Distance. Our model achieves significant and consistent improvements across all the  tasks. Best results are in bold.}
\resizebox{\linewidth}{!}{
\begin{tabular}{|cc|ccc|ccc|}
\hline
\multicolumn{1}{|c}{\multirow{2}{*}{Segmentation Task}} & \multicolumn{1}{c|}{\multirow{2}{*}{Modality}} & \multicolumn{3}{c|}{DSC(\%)}            & \multicolumn{3}{c|}{NSD(\%)}    \\ \cline{3-8} 
\multicolumn{1}{|c}{}                                   & \multicolumn{1}{c|}{}                          & SAM ~\cite{kirillov2023segment}   & MedSAM~\cite{ma2024segment}         & \textbf{DB-SAM (Ours)}           & SAM~\cite{kirillov2023segment}   & MedSAM~\cite{ma2024segment} & \textbf{DB-SAM (Ours)}           \\ \hline
Breast Tumor                                            & Ultrasound                                     & 78.01 & 85.42          & \textbf{87.43} & 82.48 & 89.02  & \textbf{92.30} \\
Liver                                                   & Ultrasound                                     & 67.81 & 74.36          & \textbf{81.10} & 72.07 & 79.07  & \textbf{92.67} \\
Vessel                                                  & Ultrasound                                     & 57.6  & 70.88          & \textbf{72.15} & 65.1  & 78.41  & \textbf{86.45} \\
Heart                                                   & X-Ray                                          & 79.28 & 91.19          & \textbf{95.35} & 83.85 & 94.1   & \textbf{95.65} \\
Lungs                                                   & X-Ray                                          & 72.24 & 96.57          & \textbf{97.60} & 75.45 & 98.56  & \textbf{99.30} \\
Polyp                                                   & Endoscope                                      & 81.6  & 86.9           & \textbf{92.13} & 85.93 & 90.91  & \textbf{96.28} \\
Instrument                                              & Endoscope                                      & 76.61 & 86.37          & \textbf{91.89} & 82.36 & 90.93  & \textbf{97.33} \\
Retinal Vessel                                          & Retinal Image                                  & 0.75  & 66.1           & \textbf{67.53} & 3.45  &84.4    & \textbf{91.36}         \\
Gland                                                   & Pathology                                      & 22.63 & 37.23          & \textbf{52.79} & 27.75 & 43.09  & \textbf{74.95} \\ \hline
Average                                                 &                                                & 59.62 & 77.22          & \textbf{82.00} & 64.27 & 83.17  & \textbf{91.81} \\ \hline
\end{tabular}
}
\label{tab:comparision on 2D tasks}
\end{table}

\subsection{Implementation and Evaluation Metrics}
\textbf{Dataset.}
Our proposed method utilizes the extensive and heterogeneous large-scale dataset collected by MedSAM \cite{ma2024segment}, adhering to their established split criteria for training and testing. The large-scale dataset encompasses 30 distinct segmentation tasks collected from public datasets across various medical devices. 
Specifically, the
segmentation tasks of  brain ventricle, brain tumor, cerebellum, gallbladder, and the left and right ventricles of the heart are generated from various MR sequences such as T1, T2, ADC, and FLAIR \cite{jia2020brats-2nd,PROMISE12,ACDC,bakas2018brats,nci_prostate_dataset,ABCs-MICCAI20,AMOS22,MMs-2020-1st,MSD-Dataset}. The segmentation tasks of abdominal tumors, COVID-19 related infections, gallbladder, head and neck tumors, liver, pancreas, pleural effusion, and stomach are generated from CT scans~\cite{Ma-2020-abdomenCT-1K,KiTS2021MIA,LiTS,FLARE21-MIA,ma2021-COVID-Data,MSD-Dataset,HECKTOR2021overview,NSCLC1,NSCLC2,TCIA}.  The segmentation tasks of heart and lungs are from X-Ray images~\cite{XRay-JSRT}.
The segmentation tasks of polyps and instruments are from endoscopic images~\cite{polyp-nature-data,Endo-instrument}.
The segmentation task of vessels is from retinal images~\cite{DRIVE}.
The segmentation task of colon glands  is from pathology images~\cite{gland-patho1,gland-patho2}. 
The images of all segmentation tasks are normalized to the range of $\left[0, 255\right]$ and resized to the fixed resolution of $256\times256\times3$. For 3D images, such as those from CT and MR scans, the images are split into 2D slices along  axial plane. Finally, the training set contains 161,857 images, while the test set comprises of 52,506 images.

\noindent\textbf{Training Settings.}
We leverage a pre-trained SAM with ViT-B and integrate our proposed adapter modules. Both the ViT encoder and the prompt encoder are kept frozen during training. Similar to MedSAM \cite{ma2024segment}, the mask decoder is trainable. During training, to closely replicate a real-world clinical setting where a doctor manually marks a target area, we generated bounding box prompts from ground-truth masks with a random perturbation ranging from 0 to 20 pixels. This approach is designed to reflect the variability and imprecision inherent in human-drawn bounding boxes in medical imaging. We train our method using 4 Nvidia A100 GPUs, and adopt  the AdamW optimizer with a polynomial learning rate. There are totally 12 epochs with  initial learning rate of 1e-4.  The adapter module is regulated by dropout and drop path with the rates of 0.4. 
We employ the sum of cross-entropy loss and dice loss to supervise  mask learning.

\noindent\textbf{Evaluation Metrics.}
Similar to MedSAM, the performance is evaluated using the standard Dice Similarity Coefficient (DSC) and Normalized Surface Distance (NSD) with 1 $mm$ tolerance, a widely recognized and reliable metrics for segmentation efficacy.

\begin{figure*}[t]
  \centering
  \includegraphics[width=\linewidth]{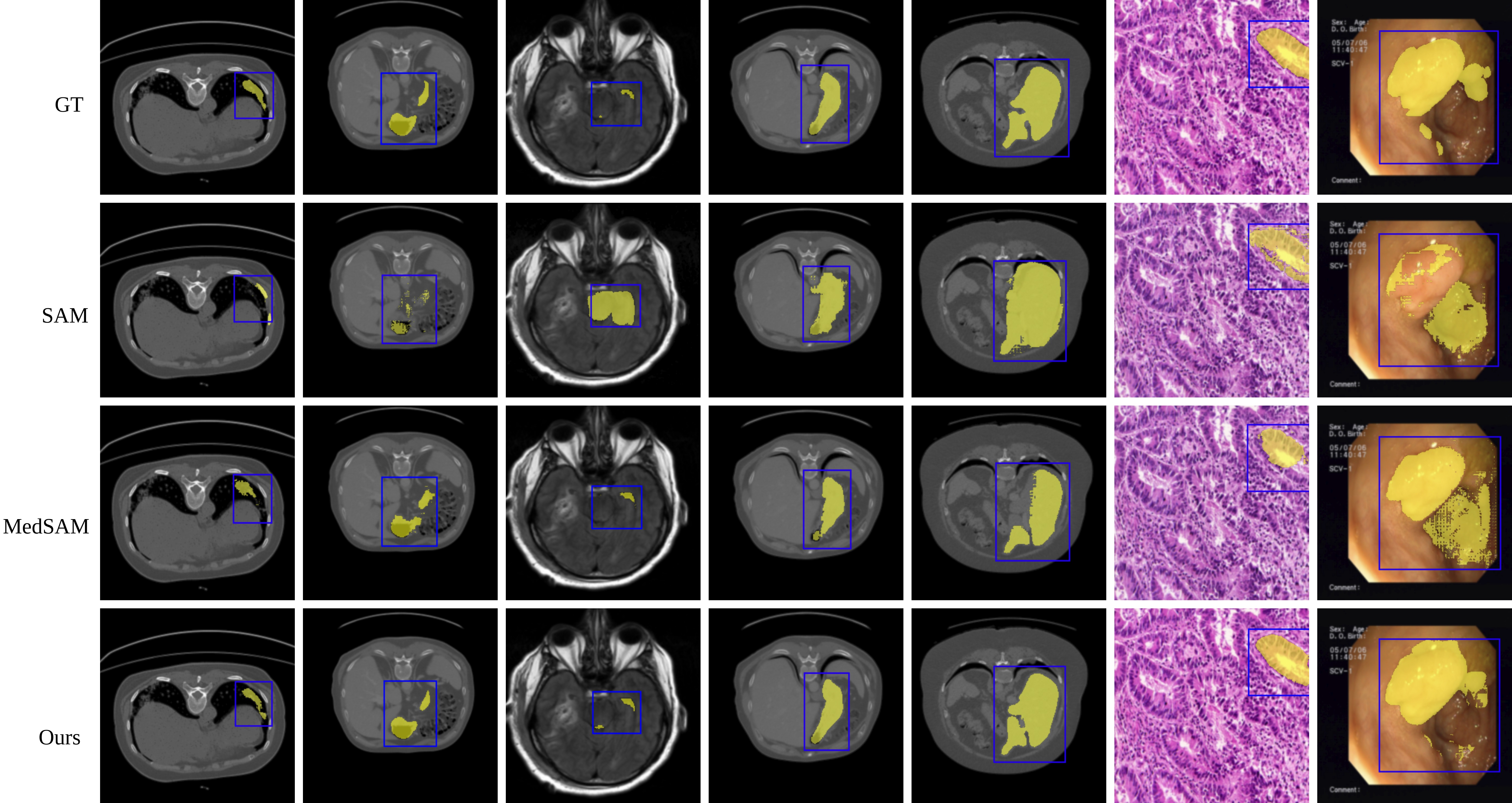}
  \caption{Visualization examples of the pre-trained SAM, MedSAM, our model and GT on different 3D and 2D tasks. Our DB-SAM model achieves more accurate segmentation than the SAM and MedSAM, especially in scenarios involving small organs and organs with complex shapes. Best viewed zoomed in. Additional results are presented in the supplementary material.}
\label{fig: comparision}
\end{figure*}

\subsection{Comparison with SAM and MedSAM}
Here we compare our DB-SAM with SAM~\cite{kirillov2023segment} and MedSAM~\cite{ma2024segment} on both 3D and 2D tasks in Table \ref{tab:comparision on 3D tasks} and Table \ref{tab:comparision on 2D tasks}. Our DB-SAM significantly outperforms both the original SAM and MedSAM across all 30 segmentation tasks.  
Compared to SAM on different 3D medical segmentation tasks, our DB-SAM achieves  28.53\% improvement in terms of average DSC and 47.82\% in terms of average NSD. Compared to MedSAM, our DB-SAM obtains  6.03\% improvement in terms of average DSC and 8.77\% in terms of average NSD. 
Further, we observe that our DB-SAM particularly excels in complex 3D tasks involving abdomen tumors, liver, lungs, pleural effusion, and stomach. For instance, on abdomen tumor segmentation, our DB-SAM obtains a DSC of 78.31\%, whereas MedSAM provides a DSC of 65.54\%. This leads to an absolute improvement of 12.77\%, likely due to  the ability of DB-SAM to capture rich local information and adapt to varying scales through its convolution branch and deformable attention operation. As a result, DB-SAM effectively addresses challenges posed by small, discrete, and irregularly shaped objects.
Table \ref{tab:comparision on 2D tasks} shows results on different 2D tasks. In terms of DSC, SAM obtains 59.62\%, MedSAM achieves 77.22\%, whereas our DB-SAM obtains significantly improved performance of 82.00\%. 
Similarly, our Db-SAM outperforms SAM and MedSAM by 27.54\% and 8.64\% in terms of NSD.

\begin{table}[!t]
\centering
  \footnotesize
\caption{Ablation study of integrating different modules, including channel attention block, bilateral cross attention block, and ViT-Conv fusion block.}

\begin{tabular}{|ccc|cc|cc|}
\hline
 \multicolumn{1}{|c}{\multirow{2}{*}{\begin{tabular}[c]{@{}c@{}}Channel \\ attention\end{tabular}}} & \multicolumn{1}{c}{\multirow{2}{*}{\begin{tabular}[c]{@{}c@{}}Bilateral \\ cross attention\end{tabular}}} & \multicolumn{1}{c|}{\multirow{2}{*}{\begin{tabular}[c]{@{}c@{}}Final \\ fusion\end{tabular}}} & \multicolumn{2}{c|}{3D Average}                     & \multicolumn{2}{c|}{2D Average}                     \\ \cline{4-7} 
\multicolumn{1}{|c}{}                                                                              & \multicolumn{1}{c}{}                                                                                      & \multicolumn{1}{c|}{}                                                                         & \multicolumn{1}{c}{Dice} & \multicolumn{1}{c|}{NSD} & \multicolumn{1}{c}{Dice} & \multicolumn{1}{c|}{NSD} \\ \hline
              &    &     &81.02        &76.54      & 77.22   &83.17                          \\
\checkmark    &    &     &85.25        &82.25      & 78.26   &89.02                          \\
\checkmark    &\checkmark   &    &86.60     &85.01      &79.55     &90.06                          \\
\checkmark      &\checkmark      &\checkmark     &\textbf{87.05}                         &\textbf{85.31 }                          &\textbf{82.00}                          &\textbf{91.81 }                          \\ \hline
\end{tabular}
\label{tab:ablation 1}
\end{table}

Fig. \ref{fig: comparision} presents a qualitative comparison of pre-trained SAM, MedSAM, and our DB-SAM with respect to ground-truth (GT).Compared to SAM and MedSAM, our proposed method can generate high-quality results. For instance, the first-row example highlights our proposed method has a superior performance in segmenting small objects with complicated shapes. Specifically, the pre-trained SAM could identify a small portion of the object, MedSAM is able to recognize some more parts. In contrast, our proposed method can accurately predict the masks of the small object. The third-row example shows that our proposed method is able to segment very small object accurately.

\subsection{Ablation Study}
\textbf{Efficacy of proposed modules}. We conduct an ablation study on impact of  three different experiments in Table \ref{tab:ablation 1}. The baseline method is MedSAM, which only fine-tunes mask decoder during training. MedSAM has an averaged DSC score of 81.02\% and an averaged NSD score of 76.54\% on 3D tasks. When we only integrate the channel attention mechanism into the ViT branch, it has an averaged DSC score of 85.25\% and an averaged NSD score of 82.25\% on 3D tasks. It outperforms the baseline by 4.21\% in terms of DSC and 5.71\% in terms of NSD. When further integrating convolution branch, it has an averaged DSC score of 86.60\% and an averaged NSD score of 85.01\% on 3D tasks. It outperforms the baseline by 5.58\% in terms of DSC and 8.47\% in terms of NSD. Finally, when fusing the output features of two branches, it has an averaged DSC score of 87.05\% and an averaged NSD score of 85.31\% on 3D tasks, which outperforms the baseline by 6.01\% and 8.77\%. Similarly, we observe that integrating different modules can improve the performance on 2D tasks.

\section{Conclusion}
We introduce a dual-branch  method, DB-SAM, to  adapt SAM  for universal medical image segmentation tasks. Our proposed DB-SAM replaces the original image encoder of SAM with a dual-branch image encoder.  The dual-branch image encoder comprises a ViT branch and a convolution branch, which are respectively used to extract high-and low-level domain-specific features. Experiments on large-scale medical image segmentation dataset demonstrate the effectiveness of proposed method, leading to superior segmentation performance.

\begin{credits}
\subsubsection{\discintname}
The authors have no competing interests to declare that are
relevant to the content of this article.
\end{credits}
%
%
%
\bibliographystyle{splncs04}
\bibliography{Paper-1489}

\begin{thebibliography}{10}
\providecommand{\url}[1]{\texttt{#1}}
\providecommand{\urlprefix}{URL }
\providecommand{\doi}[1]{https://doi.org/#1}

\bibitem{DRIVE}
Drive: Digital retinal images for vessel extraction (2023), \url{https://drive.grand-challenge.org}

\bibitem{polyp-nature-data}
Ali, S., Jha, D., Ghatwary, N., et~al.: A multi-centre polyp detection and segmentation dataset for generalisability assessment. Scientific Data  \textbf{10}(1), ~75 (2023)

\bibitem{HECKTOR2021overview}
Andrearczyk, V., Oreiller, V., Jreige, M., et~al.: Overview of the hecktor challenge at miccai 2020: Automatic head and neck tumor segmentation in pet/ct. In: Head and Neck Tumor Segmentation. pp. 1--21 (2021)

\bibitem{bakas2018brats}
Bakas, S., Reyes, M., Jakab, A., et~al.: Identifying the best machine learning algorithms for brain tumor segmentation, progression assessment, and overall survival prediction in the brats challenge. arXiv:1811.02629  (2018)

\bibitem{ACDC}
Bernard, O., Lalande, A., Zotti, C., et~al.: Deep learning techniques for automatic mri cardiac multi-structures segmentation and diagnosis: is the problem solved? IEEE Transactions on Medical Imaging  \textbf{37}(11),  2514--2525 (2018)

\bibitem{LiTS}
Bilic, P., Christ, P., Li, H.B., et~al.: The liver tumor segmentation benchmark (lits). Medical Image Analysis  \textbf{84},  102680 (2023)

\bibitem{nci_prostate_dataset}
Bloch, N., Madabhushi, A., Huisman, H., et~al.: Nci-isbi 2013 challenge: automated segmentation of prostate structures (2015), \url{https://wiki.cancerimagingarchive.net}

\bibitem{swinunet}
Cao, H., Wang, Y., Chen, J., et~al.: Swin-unet: Unet-like pure transformer for medical image segmentation. In: ECCVW (2022)

\bibitem{chen2021transunet}
Chen, J., Lu, Y., Yu, Q., et~al.: Transunet: Transformers make strong encoders for medical image segmentation. arXiv:2102.04306  (2021)

\bibitem{chen2022vision}
Chen, Z., Duan, Y., Wang, W., et~al.: Vision transformer adapter for dense predictions. arXiv:2205.08534  (2022)

\bibitem{TCIA}
Clark, K., Vendt, B., Smith, K., et~al.: The cancer imaging archive (tcia): maintaining and operating a public information repository. Journal of Digital Imaging  \textbf{26}(6),  1045--1057 (2013)

\bibitem{MMs-2020-1st}
Full, P.M., Isensee, F., J{\"a}ger, P.F., Maier-Hein, K.: Studying robustness of semantic segmentation under domain shift in cardiac mri. In: Statistical Atlases and Computational Models of the Heart. M{\&}Ms and EMIDEC Challenges (2021)

\bibitem{Endo-instrument}
Garcia-Peraza-Herrera, L.C., Fidon, L., D’Ettorre, C., et~al.: Image compositing for segmentation of surgical tools without manual annotations. IEEE Transactions on Medical Imaging  \textbf{40}(5),  1450--1460 (2021)

\bibitem{KiTS2021MIA}
Heller, N., Isensee, F., Maier-Hein, K.H., et~al.: The state of the art in kidney and kidney tumor segmentation in contrast-enhanced ct imaging: Results of the kits19 challenge. Medical Image Analysis  \textbf{67},  101821 (2021)

\bibitem{hu2018squeeze}
Hu, J., Shen, L., Sun, G.: Squeeze-and-excitation networks. In: Proceedings of the IEEE conference on computer vision and pattern recognition. pp. 7132--7141 (2018)

\bibitem{AMOS22}
Ji, Y., Bai, H., Yang, J., et~al.: Amos: A large-scale abdominal multi-organ benchmark for versatile medical image segmentation. arXiv:2206.08023  (2022)

\bibitem{jia2020brats-2nd}
Jia, H., Cai, W., Huang, H., Xia, Y.: H2nf-net for brain tumor segmentation using multimodal mr imaging: 2nd place solution to brats challenge 2020 segmentation task. In: Brainlesion: Glioma, Multiple Sclerosis, Stroke and Traumatic Brain Injuries. pp. 58--68. Springer International Publishing, Cham (2021)

\bibitem{kirillov2023segment}
Kirillov, A., Mintun, E., Ravi, N., et~al.: Segment anything. arXiv:2304.02643  (2023)

\bibitem{NSCLC2}
Kiser, K.J., Barman, A., Stieb, S., et~al.: Novel autosegmentation spatial similarity metrics capture the time required to correct segmentations better than traditional metrics in a thoracic cavity segmentation workflow. Journal of Digital Imaging  \textbf{34},  541--553 (2021)

\bibitem{NSCLC1}
Kiser, K., Ahmed, S., Stieb, S., et~al.: Data from the thoracic volume and pleural effusion segmentations in diseased lungs for benchmarking chest ct processing pipelines. The Cancer Imaging Archive  (2020)

\bibitem{PROMISE12}
Litjens, G., Toth, R., van~de Ven, W., et~al.: Evaluation of prostate segmentation algorithms for mri: the promise12 challenge. Medical Image Analysis  \textbf{18}(2),  359--373 (2014)

\bibitem{ma2024segment}
Ma, J., He, Y., Li, F., Han, L., You, C., Wang, B.: Segment anything in medical images. Nature Communications  \textbf{15}(1), ~654 (2024)

\bibitem{ma2021-COVID-Data}
Ma, J., Wang, Y., An, X., et~al.: Towards data‐efficient learning: A benchmark for covid‐19 ct lung and infection segmentation. Medical Physics  \textbf{48}(3),  1197--1210 (2021)

\bibitem{Ma-2020-abdomenCT-1K}
Ma, J., Zhang, Y., Gu, S., et~al.: Abdomenct-1k: Is abdominal organ segmentation a solved problem? IEEE Transactions on Pattern Analysis and Machine Intelligence  \textbf{44}(10),  6695--6714 (2022)

\bibitem{FLARE21-MIA}
Ma, J., Zhang, Y., Gu, S., et~al.: Fast and low-gpu-memory abdomen ct organ segmentation: The flare challenge. Medical Image Analysis  \textbf{82},  102616 (2022)

\bibitem{XRay-JSRT}
Shiraishi, J., Katsuragawa, S., Ikezoe, J., et~al.: Development of a digital image database for chest radiographs with and without a lung nodule: receiver operating characteristic analysis of radiologists' detection of pulmonary nodules. American Journal of Roentgenology  \textbf{174}(1),  71--74 (2000)

\bibitem{ABCs-MICCAI20}
Shusharina, N., Bortfeld, T., Cardenas, C., Yang, J.: Anatomical brain barriers to cancer spread: Segmentation from ct and mr images (2020), \url{https://doi.org/10.5281/zenodo.3746561}

\bibitem{MSD-Dataset}
Simpson, A.L., Antonelli, M., Bakas, S., et~al.: A large annotated medical image dataset for the development and evaluation of segmentation algorithms. arXiv:1902.09063  (2019)

\bibitem{gland-patho1}
Sirinukunwattana, K., Pluim, J.P., Chen, H., et~al.: Gland segmentation in colon histology images: The glas challenge contest. Medical Image Analysis  \textbf{35},  489--502 (2017)

\bibitem{gland-patho2}
Sirinukunwattana, K., Snead, D.R., Rajpoot, N.M.: A stochastic polygons model for glandular structures in colon histology images. IEEE Transactions on Medical Imaging  \textbf{34}(11),  2366--2378 (2015)

\bibitem{tan2021efficientnetv2}
Tan, M., Le, Q.: Efficientnetv2: Smaller models and faster training. In: International conference on machine learning. pp. 10096--10106 (2021)

\bibitem{zhou2018unet++}
Zhou, Z., Rahman~Siddiquee, M.M., et~al.: Unet++: A nested u-net architecture for medical image segmentation. arXiv:1807.10165  (2018)

\bibitem{zhu2020deformable}
Zhu, X., Su, W., Lu, L., et~al.: Deformable detr: Deformable transformers for end-to-end object detection. arXiv:2010.04159  (2020)

\end{thebibliography}

\end{document}